\documentclass[12pt]{iopart}
\usepackage{graphics}
\usepackage{rotating}
\usepackage{epsfig}
\usepackage{color}

\begin{document}

\title{Exploring quantum criticality based on ultracold atoms in optical lattices}

\author{Xibo Zhang, Chen-Lung Hung, Shih-Kuang Tung, Nathan Gemelke* and Cheng Chin }

\address{The James Franck Institute and Department of Physics, University of Chicago, Chicago, IL 60637\\
$*$ Current address: Department of Physics, The Pennsylvania State University, 104 Davey Lab, University Park, Pennsylvania 16802}
\ead{cchin@uchicago.edu}
\begin{abstract}
Critical behavior developed near a quantum phase transition, interesting in its own right, offers exciting opportunities to explore the universality of strongly-correlated systems near the ground state. Cold atoms in optical lattices, in particular, represent a paradigmatic system, for which the quantum phase transition between the superfluid and Mott insulator states can be externally induced by tuning the microscopic parameters. In this paper, we describe our approach to study quantum criticality of cesium atoms in a two-dimensional lattice based on \textit{in situ} density measurements.
Our research agenda involves testing critical scaling of thermodynamic observables and extracting transport properties in the quantum critical regime. We present and discuss experimental progress on both fronts. In particular, the thermodynamic measurement suggests that the equation of state near the critical point follows the predicted scaling law at low temperatures.
\end{abstract}


\maketitle

\section{Introduction}
\label{Intro}

A quantum phase transition occurs in a many-body system at absolute zero temperature when a new order smoothly emerges in its ground state \cite{Sachdev, Coleman}. Even though a quantum phase transition happens at zero temperature, the transition influences the finite-temperature properties of the many-body system, which leads to possible experimental observations. Near a quantum phase transition, the many-body system shows universal scaling behaviors, termed as quantum criticality, with characteristic scaling exponents determined by basic properties of the system, such as symmetry and dimensionality \cite{Sachdev}. Quantum criticality is of wide research interest because both spatial and temporal correlations in such systems are expected to follow intriguing universal relations. Understanding quantum criticality represents a major challenge to our knowledge of many-body physics.

\begin{figure}[htbp]
\begin{center}
\includegraphics [width=0.75\columnwidth,keepaspectratio]{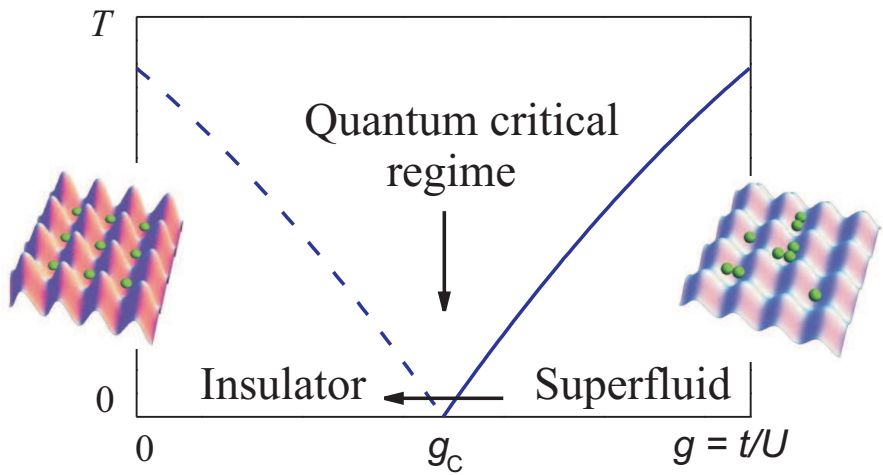}
\caption{Illustration of the phase diagram of the superfluid-to-Mott insulator quantum phase transition. The phase diagram is shown as a function of the temperature $T$ and coupling constant $g=t/U$, where $t$ is the tunneling energy and $U$ is the on-site interaction. At zero temperature, the phase transition happens at the critical point $(T=0,g=g_{\mathrm{c}})$; at finite temperatures, the system shows universal behaviors in the V-shaped quantum critical regime. The two graphs of atomic distributions in optical lattices illustrate the Mott insulator state (left) and the superfluid state (right). Our methods to probe quantum criticality are based on studying the universal scaling behaviors by changing the temperature of the system, indicated by the vertical arrow, and on studying quantum critical dynamics by ramping $g$ across the quantum phase transition, indicated by the horizontal arrow.}
\label{qcintro}
\end{center}
\end{figure}

In contrast to classical phase transitions driven by the competition between energy and entropy of a system, quantum phase transitions are driven by the competition between different quantum energy scales \cite{Sachdev, Jaksch, IB1}. In a system of bosonic atoms confined in optical lattices, the superfluid-to-Mott insulator transition, described by the Bose-Hubbard model \cite{Jaksch}, is ideal for studying quantum criticality. In this system, the two competing energy scales are the tunneling energy $t$ and the on-site interaction $U$. At zero temperature, two phases are predicted: the superfluid and Mott insulator phases, as shown in Fig. \ref{qcintro}. In the tunneling-dominated regime ($ g \equiv t/U \gg 1$), the system forms a superfluid at low temperatures; in the interaction-dominated regime ($g\ll 1$), the system assumes a Mott insulator phase in the ground state. When $g$ equals the critical value $g_c$, the ground state of the system is neither a superfluid nor a Mott insulator. At finite temperatures, the critical point expands into a V-shaped regime where universal scaling behaviors are expected \cite{Sachdev,Sachdev2, Qi09prl}, see Fig.\ref{qcintro}.

The flexible control of cold atoms and the lattice potential provide essential tools to study quantum phase transitions and quantum criticality. In optical lattices, the two competing energy scales, tunneling and on-site interaction energies, can be controlled precisely by tuning parameters of optical lattices or the magnetic field near a Feshbach resonance \cite{Chin10}. Using evaporative cooling \cite{CL, Sherson10} and other techniques \cite{Grimm, Bakr10}, one can reach a temperature low enough that the thermal fluctuations are largely suppressed. Furthermore, high resolution imaging techniques have been recently developed to measure the density profiles and fluctuations of the trapped atomic gas \cite{Sherson10, Bakr10, Nate1}. 

\begin{figure}[htbp]
\begin{center}
\includegraphics [width=0.75\columnwidth,keepaspectratio]{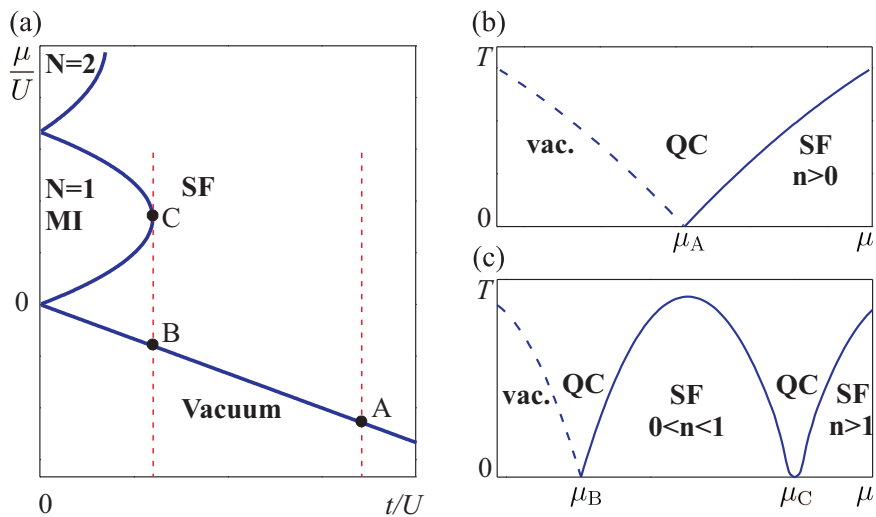}
\caption{Illustrations of phase diagrams of the Bose-Hubbard model. (a) Zero-temperature phase diagram in the $\mu/U$--$t/U$ plane. Here MI denotes the Mott insulator state, SF denotes the superfluid state, and vacuum denotes the state with no atoms occupying the lattice sites. Density profiles of trapped gases probe lines in the phase diagram along the chemical potential $\mu$ direction (with fixed $t/U$), shown by the dashed lines. Points A, B, and C are the specific critical points on the phase boundaries referred to in the text. At finite temperatures, these critical points expand to quantum critical regions. (b) Finite-temperature phase diagram along the fixed-$t/U$ line going through the vacuum-to-superfluid transition point A at zero temperature. Here vac. denotes the vacuum state and QC denotes the quantum critical region. (c) Finite-temperature phase diagram along the fixed-$t/U$ line going through the vacuum-to-superfluid transition point B and the tip of the occupation number $N = 1$ Mott insulator lobe (point C) at zero temperature.}
\label{BHM}
\end{center}
\end{figure}

In addition to the optical lattice, atoms are typically confined by an external harmonic potential, which adds new perspectives for experimental observations. Because of this potential, the system has a higher density as one moves toward the trap center \cite{Jaksch, Nate1}. This density profile can reveal the phase diagram of a homogeneous system with different chemical potentials at a fixed coupling constant $g$, as shown in Fig. \ref{BHM}(a). A system with a large and negative chemical potential is in a vacuum state. As one increases the chemical potential, the system evolves into the superfluid state. For small enough $g$ and larger chemical potential, the system can enter the Mott insulator state with unit occupation number.  Typically, the harmonic potential is slowly varying compared to the optical lattice potential, and the density distribution is expected to be slowly varying over the length scale of the lattice constant. With a weak harmonic confinement, the local density approximation applies, and every point in the system can be viewed as a homogeneous sub-system with a local chemical potential. Therefore, a measured density profile probes a line in the phase diagram along the chemical potential direction. Near the phase boundary, the line crosses the quantum critical regime of our research interest, as shown in Fig. \ref{BHM}. 

The main goal of this paper is to describe two possible ways of observing quantum criticality in optical lattices. One is through the analysis of equilibrium atomic density profiles, and the other is through dynamics of the sample in the quantum critical regime. We start in Sec.~\ref{theory} with a brief review of the theoretical basis of critical universality in a system of ultracold atoms in optical lattices. In Sec.~\ref{Exp}, we present progress toward observing this critical regime from analyses of \textit{in situ} atomic density profiles. Also, our observations on the mass flow in optical lattices demonstrate the possibility of observing dynamics in the quantum critical regime.

\section{Probing quantum criticality in the density profiles and dynamics of ultracold atoms}\label{theory}

We explore quantum critical behaviors of ultracold atoms near the superfluid-to-Mott insulator transition in a 2D optical lattice. Based on high-resolution \textit{in situ} absorption imaging, we can record density profiles and fluctuations of the sample in equilibrium, which allows us to identify critical scaling laws of thermodynamic observables and compare to the theoretical predictions. Another direction is to initiate a dynamic passage to the quantum critical regime, from which we attempt to extract dynamical critical exponents and quantum transport coefficients. In this section, we will discuss  the ``scaling" and the ``dynamics" approaches separately.

\subsection{Scaling of equation of state}
Sufficiently near the quantum critical point, thermodynamic quantities, such as the atomic density $n$, obey universal scaling relations \cite{FisherT2, QiQC, KadenQC, DWQC}. Consider the equation of state $n = n(\mu,T)$, where $n$ is the density, $\mu$ is the chemical potential, and $T$ is the temperature. Near the superfluid-to-Mott insulator phase boundary, finite temperature leads to excitations in the Mott-insulating domain, and the occupation number $nd^2$ deviates from the non-negative integer occupation number $n_0 d^2 = 0, 1, 2, \cdots $ in the zero-temperature limit. Here $d$ is the optical lattice constant, and the $n_0 d^2 = 0$ state corresponds to the vacuum state. After the non-universal part $n_0$ is subtracted from $n$, the remaining part $n_u$  is expected to satisfy the following scaling form  in the critical regime\cite{FisherT2, QiQC, KadenQC}
\begin{eqnarray}\label{scalingFormula}
n_u(\mu,T) &  =  & n(\mu,T) - n_0(\mu,T) \nonumber\\
& = & d^{-D} \left(\frac{k_{\mathrm{B}}T}{t}\right)^{\frac{D}{z}+1-\frac{1}{z \nu}}F\left(\frac{\frac{\mu - \mu_{\mathrm{c}}}{t}}{(\frac{k_{\mathrm{B}}T}{t})^{\frac{1}{z \nu}}}\right),
\end{eqnarray}
where $D$ is the dimensionality, $F$ is a generic function, $z$ is the dynamical exponent, $\nu$ is the correlation length exponent, $k_{\mathrm{B}}$ is the Boltzmann constant, $\mu_{\mathrm{c}}$ is the chemical potential where the zero-temperature phase transition takes place. The scaling form Eq.~(\ref{scalingFormula}) holds when the dimensionality $D$ is below the upper critical dimension $D_{\mathrm{c}} = 4-z$, where $z$ will be $1$ or $2$ depending on what part of the phase boundary the critical point resides at. Near the generic vacuum-to-superfluid transition points (see points A and B in Fig. \ref{BHM}(a)) and the superfluid-to-Mott insulator transition points (other than the tip of the Mott insulator lobe), the equations of state are expected to scale according to the dilute Bose gas universality class with critical exponents $z=2,\nu=1/2$; at the tip of the Mott insulator lobe (see point C in Fig. \ref{BHM}(a)), the equation of state scales according to the $O(2)$ rotor universality class with critical exponents $z=1,\nu=1$ \cite{Sachdev, FisherT2, KadenQC, DWQC}.  

We study the critical scaling relations based on \textit{in situ} absorption imaging of atomic density profiles. Applying the local density approximation (LDA) enables deduction of the properties of homogeneous systems from the observed properties of non-uniform trapped atoms \cite{QiQC, scalettar09pra}. Each density profile corresponds to a line along the chemical potential direction in the phase diagram (for example, see the two red dashed lines in Fig. \ref{BHM}(a)). Sufficiently close to the quantum critical point, when the scaled equation of state $\frac{(n - n_0)d^D}{(k_{\mathrm{B}}T/t)^{\frac{D}{z}+1-\frac{1}{z\nu}}}$ is plotted versus the chemical potential $\mu$ for different temperatures,  all curves should intersect at the same point $\mu = \mu_{\mathrm{c}}$. Once $\mu_{\mathrm{c}}$ is determined, one can plot the scaled equation of state versus the scaled chemical potential $\frac{(\mu - \mu_{\mathrm{c}})/t}{(k_\mathrm{B}T/t)^{\frac{1}{z\nu}}}$, and all curves for different temperatures should collapse into a single curve given by the generic function $F$ \cite{QiQC}.

Finally, we point out important issues on observing quantum criticality in optical lattice experiments. First of all, since quantum critical scaling laws are valid only at sufficiently low temperatures, it remains an interesting question to determine and understand the deviation from the expected scaling laws due to  finite-temperature effects \cite{Qi09prl,DWQC}. Recent Quantum Monte Carlo simulations suggest that universal scaling laws can be observed at temperatures as high as $k_{\mathrm{B}}T \sim 6t$ for atoms in a two-dimensional lattice, where the width of the ground band is $8t$ \cite{DWQC}. Secondly, inhomogeneity of the trapping potential can suppress the quantum critical behavior at low temperatures when the correlation length is restricted by the length scale of the trapping potential \cite{trapPRA04}.  The finite-temperature and trap effects require the temperature to be small compared with the ground band width and large compared with the characteristic energy scale of the trapping potential. Finite size scaling algorithms, which have been applied to improve the accuracy of analysis in quantum Monte Carlo simulations \cite{Barbara08pra}, have also been proposed for the experiments \cite{QiQC}.

\subsection{Quantum critical dynamics}\label{theoryDynamics}
We discuss some prospects of studying quantum critical dynamics using cold atoms in optical lattices. Prominent dynamic phenomena include quantum critical transport of mass and entropy, and dynamics of defect generation across the quantum critical point as described by the Kibble-Zurek mechanism \cite{Zurek05,Cucchietti07}.

\noindent \textit{A. Quantum critical transport}

Mass and heat transport across the quantum critical regime provide important tests for quantum critical theory \cite{Sachdev}. Sufficiently close to the critical point, one expects that transport coefficients obey universal scaling relations independent of microscopic physics \cite{Sachdev, Sachdev00}. In two dimensions, in particular, we expect that the static mass transport exhibits a universal behavior, in analogous to the prediction on the electrical conductivity \cite{Cha91,Sachdev09}, and the static mass conductivity at the critical point is given by
\begin{equation}\label{staticMassCond}
\sigma  = \frac{m}{\hbar} \Phi_{\sigma},
\end{equation}
which only depends on the fundamental constants $m/\hbar$ and a dimensionless, universal number $\Phi_\sigma$ determined from the universality class of the underlying phase transition. Here $\hbar$ is the reduced Planck constant, and $m$ is the atomic mass. Analytic predictions on the transport coefficients in the quantum critical regime were recently reported on the basis of the anti-de Sitter/conformal field theory duality \cite{Sachdev09,AdSCFT}. Measurements of transport coefficients in general can be of fundamental interest in quantum field theory \cite{Sachdev09}; the relation between mass and thermal conductivities is in close analogy to the Wiedemann-Franz relation between charge and thermal transport coefficients in electronic systems, which is shown to break down near the quantum critical point in a recent experiment \cite{Tanatar07}.

Mass and heat transport are induced by generalized forces such as chemical potential gradient and temperature gradient.  A natural approach to study dynamics of atoms in optical lattices is to first create non-equilibrium density distributions in the sample and then measure the subsequent evolution of density profiles.

\begin{figure}[htbp]
\begin{center}
\includegraphics [width=0.8\columnwidth,keepaspectratio]{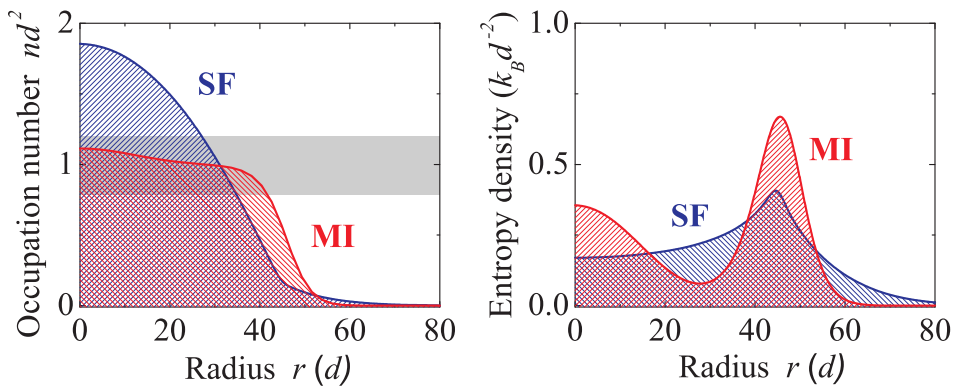}
\end{center}
\caption{Sketch of density and entropy profiles of a trapped, finite-temperature gas in the superfluid regime (SF, blue curves) and the Mott-insulating regime (MI, red curves), with the same particle number and total entropy. The gray shaded area marks an extended region near unit site occupation number ($nd^2=1$) where quantum critical transport can take place when global adiabaticity breaks down during the lattice loading process. $d=532$~nm is the lattice constant.}
\label{density_and_entroy}
\end{figure}

Non-equilibrium density distributions can be induced in various ways. For example, one can create a controlled perturbation in the local chemical potential and induce transport by dynamically changing the envelope trapping potential in an equilibrated system or changing the on-site interaction $U$ near a Feshbach resonance \cite{Chin10}. On the other hand, applying lattice ramps slow compared to local microscopic time scales can still violate global adiabaticity and induce macroscopic mass and heat flow \cite{Hung10}. This is aggravated by the pronounced difference in the equilibrium density and entropy profiles between superfluid and Mott insulator phases, as shown in Fig.~\ref{density_and_entroy}. In a non-equilibrated system, we expect quantum critical dynamics to take place near integer site occupation numbers.

While measuring the evolution of the density profile is straightforward using our \textit{in situ} imaging technique \cite{Nate1, Hung10}, heat or entropy measurement in the quantum critical regime remains a challenging task. Nevertheless, the entropy profile is readily measurable deeply in the Mott-insulating regime by counting occupancy statistics using single-site resolved florescence imaging in combination with on-site number filtering \cite{Sherson10, Bakr10}, or can be extracted from counting average site occupancies before and after on-site number filtering processes \cite{Hung10}. Since the local equilibration time scale (on the order of $\hbar/U$ \cite{Bakr10}) is sufficiently decoupled from the global dynamics \cite{Stefan10}, a locally isentropic projection from the quantum critical regime deeply into the MI regime can be achieved and the local entropy profile measured.

From the density and entropy profile measurements, we can determine their current densities through the application of a generic continuity equation $\frac{\partial \rho }{\partial \tau }+ \nabla \cdotp \vec{J_{\rho}} = \Gamma_{\rho}$. Here $\rho(\vec{x},\tau)$ represents experimentally measured mass or entropy density, $\vec{J_{\rho}}$ is the corresponding current density, and $\Gamma_{\rho}$ is a source term which characterizes, for example, particle loss ($\Gamma_{\rho} < 0$) or entropy generation ($\Gamma_{\rho} > 0$).

Mass conductivity $\sigma$ and thermal conductivity $\kappa$ can be determined by relating the mass and entropy current densities, $\vec{J}_n$ and $\vec{J}_s$, as functions of position $\vec{x}$ and time $\tau$, to the generalized forces: the local chemical potential gradient $\vec{\nabla} \mu$, the potential energy gradient $\vec{\nabla} V$, and the temperature gradient $\vec{\nabla} T$. They obey the following transport equations \cite{Rammer}
\begin{eqnarray}
 \vec{J_n}(\vec{x},\tau) &=& - \sigma \vec{\nabla}[ \mu(\vec{x},\tau) + V(\vec{x},\tau)] - \frac{m L_{nq} }{k_B}\vec{\nabla} T (\vec{x},\tau)\label{masseq}\\
 \vec{J}_s(\vec{x},\tau) &=& - \frac{L_{qn}}{k_B} \vec{\nabla} [\mu(\vec{x},\tau) + V(\vec{x},\tau)] - \frac{\kappa}{T} \vec{\nabla} T(\vec{x},\tau).\label{entropyeq}
\end{eqnarray}
Here $L_{nq}$ and $L_{qn}$ are phenomenological coefficients similar to the Seebeck and Peltier coefficients in the thermoelectric effect and can be related via the Onsager reciprocity relation \cite{Grandy08}.

Finally, to obtain precise information of spatially resolved chemical potential gradient and temperature gradient, we resort to the equilibrium properties of the sample which can be determined from measurements of the equilibrium density and density fluctuation. The complementary knowledge of the equation of state $n(\mu,T)$ and its fluctuation $\delta n^2(\mu,T)$ in equilibrium can be inverted to obtain $\mu(n,\delta n^2)$ and $T(n,\delta n^2)$. We propose that, in a sample driven out of equilibrium globally but remaining locally equilibrated, local density and fluctuation measurements can still be used to extract its local chemical potential and temperature. This assumption can be further examined by comparing local compressibility to density fluctuation and extracting local temperature through the application of the fluctuation-dissipation theorem \cite{Zhou09}.

\noindent \textit{B. The Kibble-Zurek mechanism (KZM)}

Based on general critical scaling arguments, the KZM predicts the formation of topological defects after a system dynamically crosses through a second-order thermodynamic \cite{Kibble,Zurek} or quantum phase transition \cite{Zurek05,Cucchietti07}. For optical lattice experiments, the KZM applies when the system is quenched from a gapped Mott insulator state to a gapless superfluid phase, and predicts that the density of defects scales with the ramp rate of the coupling strength $g$ \cite{Cucchietti07}.

The scaling behavior can reveal critical exponents of the underlying quantum phase transition \cite{Cucchietti07,Polkovnikov05, Grandi10}. When the coupling strength $g$ is adiabatically ramped close to $g_c$, the many-body gap $\Delta$ scales as $\Delta \propto |g-g_c|^{z\nu}=|\lambda(\tau)|^{z\nu}$, where $\lambda(\tau) =g(\tau)-g_c$ characterizes the time dependence of the ramp. The adiabaticity criterion breaks down at a time $\tau=\tau_a$ when the gap $\Delta$ becomes small enough and the ramp rate violates $\frac{d \Delta}{d \tau} \frac{\hbar}{\Delta} \leq \Delta$, yielding excitations with a characteristic  energy scale $\Delta_a \propto |\lambda'(\tau_a)|^{z\nu/(z\nu +1)}$ or length scale $\xi_a \propto \Delta_a^{-1/z}$. The density of defects $n_{ex}$ should therefore scale universally as \cite{Cucchietti07, Polkovnikov05}
\begin{eqnarray}
n_{ex} \propto \xi_a^{-D} \propto \gamma_a^{D\nu/(z\nu+1)},
\end{eqnarray}
where $\gamma_a = |\lambda'(\tau_a)|$ is the magnitude of the ramp rate at which adiabaticity fails.

In a two-dimensional superfluid, topological defects are vortices. Observing vortices in an optical lattice using \textit{in situ} imaging is challenging, largely due to the smallness of a typical vortex core size ($<1~\mu$m) compared to the imaging resolution ($\geq 5~\mu$m) available in most experiments. While the latter can be technically improved, increasing the vortex core size by reducing the atomic interaction can also be achieved experimentally either through tuning a magnetic Feshbach resonance \cite{Chin10} or releasing atoms for a short time-of-flight time \cite{AboShaeer01, Coddington04}.

Further extensions of the KZM consider finite-temperature and finite-size effects \cite{Grandi10}. In general, the scaling of excitations also depends on the pathway of quenching \cite{Cucchietti07,Grandi10,Schutzhold06}, and the system can enter the Landau-Zener regime in nearly defect-free processes \cite{Zurek05}. Detailed experiments could reveal the wealth in the dynamics of quantum critical phenomena as well as the intriguing connection between quantum mechanics and thermodynamics in genuine quantum systems \cite{Polkovnikov08, Rigol08}.


\section{Experimental progress}\label{Exp}
\subsection{Scaling of density profiles}\label{expsca}
We study quantum critical scaling relations based on \textit{in situ} imaging of atomic density profiles. Our experimental procedure follows three steps. First, we prepare a quantum gas of $10^4$ cesium atoms in a two-dimensional (2D) trap with an aspect ratio of 200 to 1 by loading a three-dimensional Bose-Einstein condensate into a single site of a vertical lattice with a lattice constant of 4 $\mu\mathrm{m}$. Secondly, we transfer the 2D quantum gas into a 2D optical lattice (with lattice constant $d = 532 $ nm) which induces the superfluid-to-Mott insulator transition in sufficiently deep lattices. Finally, we perform high-resolution absorption imaging to record the atomic density profiles. Details on the apparatus and experimental sequence can be found in earlier reports \cite{Nate1, Hung10, CL2}.

Previously, we observed scale invariance and universality in two-dimensional Bose gases with different temperatures and interaction strengths \cite{CL2}. This observation motivates us to verify the universal scaling relations near the quantum phase transition for atoms in optical lattices. We work with scattering length $a = 300 a_{\mathrm{B}}$, lattice depth $V_{\mathrm{Lat}} = 6.8 E_{\mathrm{R}}$ where $E_{\mathrm{R}} = h \times 1\mathrm{,}326 \mathrm{Hz}$ is the recoil energy, and $h$ is Planck's constant.
At this lattice depth, the energy gap between the ground and first excited bands is much larger than the temperature of the system, and the system is governed by ground-band physics; we can reach a temperature as low as $T = 4t/k_{\mathrm{B}}$, where $t = k_{\mathrm{B}}\times 2.7 $ nK is the tunneling parameter.

Near the vacuum-to-superfluid transition point A in Fig. \ref{BHM}, the non-universal part of the density, $n_0$, introduced in Eq.~(\ref{scalingFormula}), equals zero, and the density profile $n$ is expected to obey the scaling relation of the dilute Bose gas universality class\cite{FisherT2,KadenQC}:
\begin{eqnarray}\label{VacToSF}
\frac{nd^2}{k_{\mathrm{B}}T/t} = f\left(\frac{\mu - \mu_c}{k_{\mathrm{B}}T}\right),
\end{eqnarray}
where $f$ is a universal function, the exponents are $z = 2, \nu = 1/2$. We expect the zero-temperature vacuum-to-superfluid transition in 2D lattices to happen at a critical chemical potential $\mu_{\mathrm{c}} = -4t$, which can be understood in the following picture. The ground band of a 2D lattice has a width of $8t$, with the middle of the band chosen as the chemical potential zero point. At zero temperature, when the chemical potential is below the bottom of the band ($\mu < -4t$), atoms are not allowed in the lattice and the ground state is a vacuum; as soon as the chemical potential reaches $-4t$, atoms are allowed to occupy the lowest energy state and will form a superfluid. Thus the zero-temperature phase boundary is $\mu_{\mathrm{c}} = -4t$ for the vacuum-to-superfluid transition. At sufficiently low temperatures, we expect that the constant-temperature traces of the scaled equation of state, when plotted as functions of the chemical potential $\mu$ at different temperatures, should have a common crossing point at $\mu_{\mathrm{c}}$, as is suggested by Eq.~(\ref{VacToSF}).

\begin{figure}[htbp]
\begin{center}
\includegraphics [width=0.85\columnwidth,keepaspectratio]{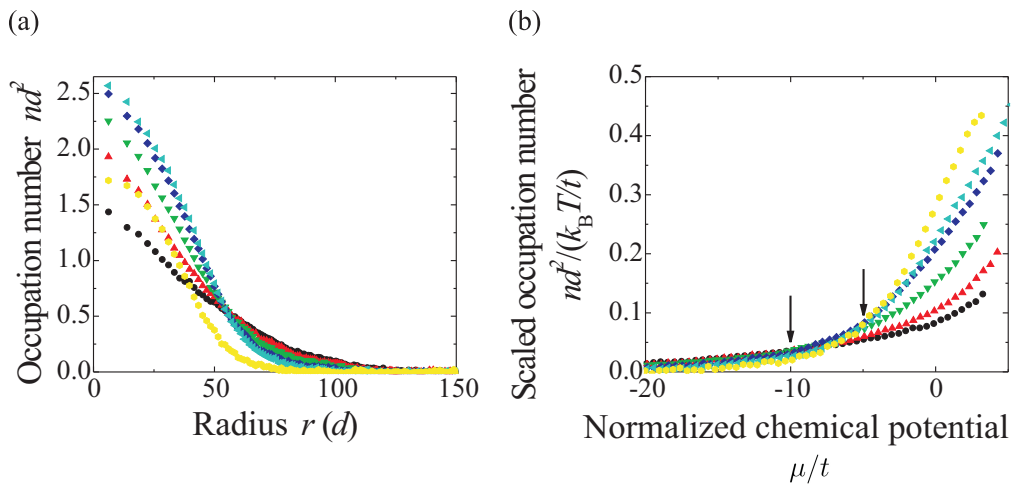}
\caption{Density profiles and the scaled equation of state at different temperatures (at $V_{\mathrm{Lat}} = 6.8 E_{\mathrm{R}}$, $t/U = 0.18$). (a) Occupation number as a function of radius measured at different temperatures: $T = $ 29 nK (black circles), 26 nK (red triangles), 24 nK (green triangles), 18 nK (blue diamonds), 15 nK (cyan triangles), and 11 nK (yellow hexagons, with a lower particle number).  (b) Scaled equation of state as a function of local chemical potential. When $k_{\mathrm{B}}T \approx 10t$, the constant-temperature traces of the scaled equation of state cross at $\mu \approx -10 t$; when $k_{\mathrm{B}}T \approx 5t$, the curves cross at $\mu \approx -5t$, shown by the two arrows. }
\label{density}
\end{center}
\end{figure}

To verify this scaling relation, we take a series of atomic density profiles at different temperatures, shown in Fig. \ref{density}(a). We extract the temperature and peak chemical potential by fitting the low-density tails according to the following formula, which is originally derived for non-interacting gases \cite{Jason10NatP} and is here modified to include the mean-field interaction effect, and to include all orders in the fugacity expansion \cite{QiTalk}:
\begin{eqnarray}\label{meanfield}
n(\vec{x}) & = & \frac{1}{d^2}\sum_{l=1}^{\infty}\left[I_0\left(2 l \beta t\right)\right]^2 e^{l\beta\left[\mu_0 - 2g_{\mathrm{eff}}n(\vec{x}) - V(\vec{x})\right]},
\end{eqnarray}
where $n$ is the 2D atomic density, $d = 0.532 \mu\mathrm{m}$ is the 2D lattice constant, $I_0(x) = \int^{\pi}_{-\pi}\frac{\mathrm{d}\theta}{2\pi} \exp\left(x\cos\theta\right)$ is the zeroth order Bessel function with purely imaginary argument, $\beta = \frac{1}{k_{\mathrm{B}}T}$, $\mu_0$ is the peak chemical potential, $g_{\mathrm{eff}} = Ud^2$ \cite{QiTalk} is the effective interaction strength for a 2D thermal gas in a 2D lattice, and $V$ is the trapping potential.

\begin{figure}[htbp]
\begin{center}
\includegraphics [width=0.9\columnwidth,keepaspectratio]{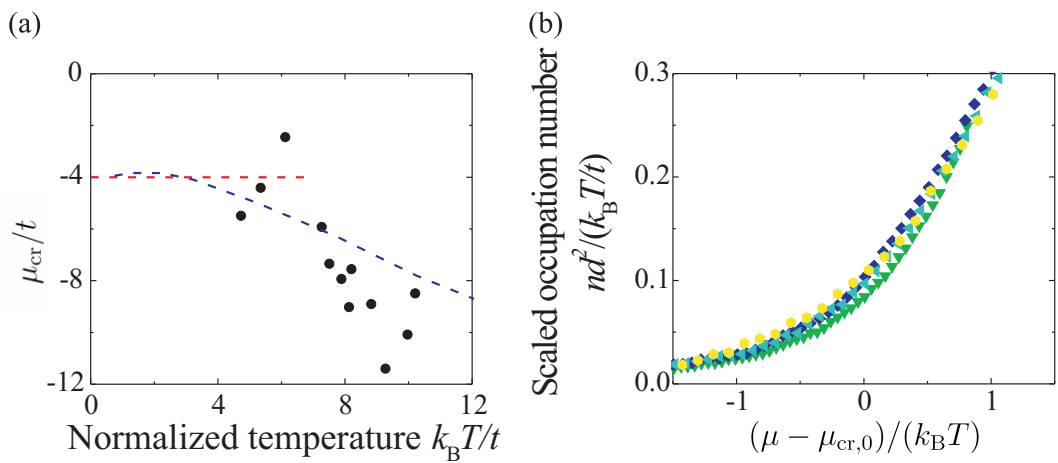}
\caption{Crossing points and overlapping of the constant-temperature traces of the scaled equation of state (at $V_{\mathrm{Lat}} = 6.8 E_{\mathrm{R}}$, $t/U = 0.18$). In (a), the chemical potential at the crossing points defined in Fig.~\ref{density} (b) is plotted as a function of the average temperature (black solid circles). The asymptote in the zero-temperature limit, $\mu_{\mathrm{cr},0} = -4t$, and the mean-field predictions based on Eq.~(\ref{meanfield}) are shown as the red dashed line and the blue dashed line, respectively. In (b), the scaled equation of state is plotted against the scaled chemical potential based on $\mu_{\mathrm{cr},0}=-4t$ for the low-temperature data: 11 nK (yellow hexagons), 15 nK (cyan triangles), 18 nK (blue diamonds), and 24 nK (green triangles). Here the ground band in a 2D optical lattice potential has a width of $8t=$21 nK.}
\label{crossing}
\end{center}
\end{figure}

By plotting the scaled equation of state as a function of the chemical potential $\mu$, shown in Fig. \ref{density}(b), we see that the constant-temperature traces of the scaled equation of state at different temperatures cross at different chemical potentials. We study this temperature dependence by finding the crossing point for each pair of constant-temperature traces of the scaled equation of state, and plot the chemical potential at the crossing point as a function of the average temperature of the pair, shown in Fig. \ref{crossing} (a). As the temperature decreases, the crossing points move toward higher chemical potentials. We take the data with the lowest four temperatures, and plot the scaled equation of state $\frac{nd^2}{k_{\mathrm{B}}T/t}$ as a function of the scaled chemical potential $\frac{\mu - \mu_{\mathrm{cr,0}}}{k_{\mathrm{B}}T}$, where $\mu_{\mathrm{cr,0}} = -4 t$ is the expected zero-temperature transition point, shown by the red dashed line in Fig. \ref{crossing} (a). Taking $\mu_{\mathrm{cr},0} = -4t$, we can test the scaling law, Eq.~(\ref{VacToSF}), by rescaling the density and chemical potential. At low temperatures (10 nK $\leq T \leq$ 24 nK), the rescaled data, shown in Fig.~\ref{crossing} (b), are much closer to each other than they are in Fig.~\ref{density} (b), which is consistent with the critical scaling law given by Eq.~(\ref{VacToSF}), near the critical point $\mu_{\mathrm{cr},0} = -4t$.

At low temperatures, our preliminary measurement on the equation of state is consistent with the expected critical scaling law given by Eq.~(\ref{VacToSF}). To test the critical scaling law more quantitatively, it is important to determine the peak chemical potential and temperature of the sample accurately in the fitting. Further analysis is underway to reduce the systematic errors in the fitting, and to test scaling laws near points B and C in the phase diagram (see Fig. \ref{BHM}).

\subsection{Dynamics}\label{expdyn}
Our recent experiment studied global mass transport and statistical evolution  in a 2D sample across the SF-MI phase boundary \cite{Hung10}. We discovered  slow equilibration dynamics with time scales more than $100$ times longer than the microscopic time scales for the on-site interaction and tunneling energy. This suggests that transport can limit the global equilibration process inside a sample traversing a quantum critical point.

\begin{figure}[htbp]
\begin{center}
\includegraphics [width=0.8\columnwidth,keepaspectratio]{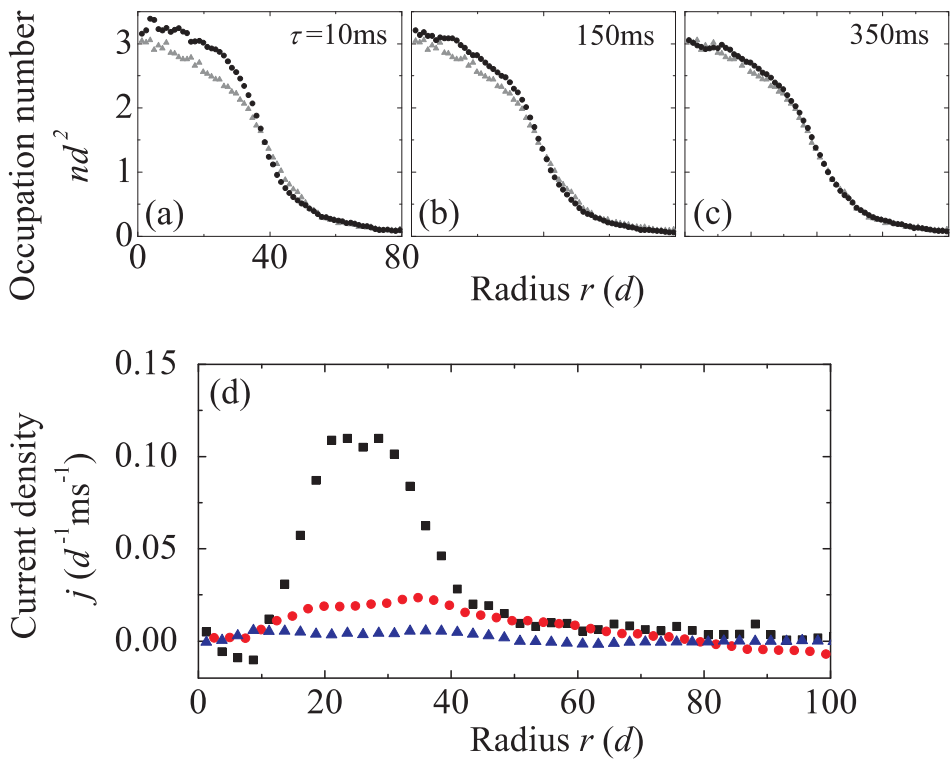}
\end{center}
\caption{Evolution of the density profile and the atom number current density after a short $50$~ms lattice ramp from zero depth to a final depth of 10~$E_R$ ($U/t=11$). Upper figure shows the density profile after holding the sample at a constant final depth for hold times $\tau=$(a) 10, (b) 150, and (c) 350~ms (black circles). In each figure (a-c), the near-equilibrated density profile measured at long hold time $\tau=500$~ms (gray triangles) is plotted for comparison. (d) shows the atom number current density at hold time $\tau=10$ (black squares), 150 (red circles), and 350 (blue triangles)~ms, derived from the density profiles measured near hold times shown in (a-c) using Eq.~(\ref{masseq}).}
\label{massflow}
\end{figure}

In Fig.~\ref{massflow}~(a-c), we plot the evolution of density profiles of a 2D gas containing $N = 2 \times 10^4$ atoms after a short $50$~ms ramp from zero to a final lattice depth of $10$~$E_R$. At this lattice depth, $U/t = 11$ is below the critical point $U/t=17$ for the Mott insulator state with unit occupation number \cite{Barbara08pra}. We record density profiles after holding the sample at the final lattice depth for various hold times $\tau$. With an equilibration time scale around $180$~ms, the cloud gently expands and the peak density slowly decreases due to the increase of repulsive atomic interaction during the lattice ramp. This equilibration time scale can depend on the sample size and the local properties of the coexisting phases in an inhomogeneous sample. 

We further extract the evolution of local mass current density, leading to detailed local transport properties beyond a single equilibration time scale. We compute the mass current density $\vec{J_n}$ by comparing density profiles taken at adjacent hold times ($\Delta \tau = 10 \sim 50$~ms) and applying the continuity equation, $m \frac{\Delta n}{\Delta \tau} + \nabla \cdotp \vec{J_n}(r,\tau) = 0$, to evaluate $\vec{J_n} (r,\tau)$. Here, we assume no atom loss in the analyses for short hold times $\tau<500$~ms. Assuming that mass flow only occurs in the radial direction ($\hat{r}$) due to azimuthal symmetry of the sample, we write the mass current density as $\vec{J_n}(r,\tau)=m j(r,\tau) \hat{r}$. The number current density $j(r,\tau)$ is computed according to
\begin{equation} 
j(r,\tau)= \frac{1}{2\pi r} \frac{ N(r,\tau+\Delta \tau) - N(r,\tau)}{\Delta \tau},
\end{equation}
where $N(r,\tau)=\int_0^r n(r',\tau) 2\pi r' \mathrm{d} r'$ is the number of atoms located inside a circle of radius $r$ at hold time $\tau$. Positive $j$ means a current flowing toward larger radius $r$, and vice versa.

In Fig.~\ref{massflow}(d), we show $j(r,\tau)$ computed from density profiles measured near hold times shown in Fig.~\ref{massflow}~(a-c). We observe overall positive mass flow, which is consistent with the picture of an expanding sample inside the optical lattice. The mass current density varies across the sample. Shortly after the lattice ramp at $\tau=10$~ms, mass transport is most apparent inside a radius $r = 40d$, where the occupation number $nd^2 >1$ and the atoms respond to the increase of on-site repulsion. The current density peaks around an annular area $20d <r <30d$ when the occupation number is in the range $2<nd^2<3$;  outside this annular area, the current density is suppressed when the occupation number is in the range $nd^2 > 3$ or $nd^2 < 2$. At a larger hold time $\tau=150$~ms, similar transport continues to take place but with smaller amplitude. At a long hold time $\tau=350$~ms when the sample is closer to equilibration, the current density $j$ becomes smaller than our measurement noise.

In this section, we have shown that spatially resolved mass current density is readily measurable using our \textit{in situ} imaging technique. We expect that local transport coefficients can be extracted using Eq.~(\ref{masseq}), from further measurements of local temperature gradients and chemical potential gradients. Our interest lies in mass transport in the quantum critical regimes near integer occupation numbers, where the static mass conductivity is predicted to be  universal (Eq.~(\ref{staticMassCond})). Measurements of local entropy density are under future investigations, with details outlined in section \ref{theoryDynamics}.

\section{Acknowledgements}
We thank S. Fang,  C.-M. Chung, D.-W. Wang, Q. Zhou, T.-L. Ho and K. Hazzard for discussions, and L.-C. Ha for careful reading of the manuscript. This work was supported by NSF (grant numbers PHY-0747907, NSF-MRSEC DMR-0213745), the Packard foundation, and a grant from the Army Research Office with funding from the DARPA OLE program. N.G. acknowledges support from the Grainger Foundation.

\end{document}